\documentclass[doublecol]{epl2} 
\usepackage{graphics}
\usepackage{epsfig}
\usepackage{amssymb}
\usepackage{subfigure}
\usepackage{pifont}

\title{Soft wetting and the Shuttleworth effect, at the crossroads between thermodynamics and mechanics}
\shorttitle{Soft wetting and the Shuttleworth effect} 

\author{Bruno Andreotti$^1$ and Jacco H. Snoeijer$^{2,3}$}
\shortauthor{B. Andreotti \& J.H. Snoeijer}
\institute{$^1$Physique et M\'ecanique des Milieux H\'et\'erog\`enes, UMR 7636 ESPCI -- CNRS -- Univ.~Paris-Diderot -- Univ.~P.M.~Curie, 10 rue Vauquelin, 75005 Paris, France.\\
$^2$Physics of Fluids Group, Faculty of Science and Technology, University of Twente, P.O. Box 217, 7500 AE Enschede, The Netherlands.\\
$^3$Department of Applied Physics, Eindhoven University of Technology, P.O. Box 513, 5600MB, Eindhoven, The Netherlands}


\abstract{
Extremely compliant elastic materials, such as thin membranes or soft gels, can be deformed when wetted by a liquid drop. It is commonly assumed that the solid capillarity in ``soft wetting" can be treated in the same manner as liquid surface tension. However, the physical chemistry of a solid interface is itself affected by any distortion with respect to the elastic reference state. This gives rise to phenomena that have no counterpart in liquids: the mechanical surface stress is different from the excess free energy in surface. Here we point out some striking consequences of this ``Shuttleworth effect" in the context of wetting on deformable substrates, such as the appearance of elastic singularities and unconventional capillary forces. We provide a synthesis between different viewpoints on soft wetting (microscopic and macroscopic, mechanics and thermodynamics), and point out key open issues in the field.}

\begin{document}

\maketitle

The canonical example of elasto-capillarity consists of a liquid drop in contact with a highly  deformable elastic material \cite{RomanJPCM}. The forces of surface tension of the liquid can induce wrinkles on a thin membrane~\cite{Huang2007,SchrollPRL13,HurePRL}, bundling of slender rods \cite{BicoNATURE,BoudPRE,ChiodiEPL}, capillary origami~\cite{PRDBRB07,PyEPJST,HonsAPL,ReisSM10}, and the slowing down of droplets moving over soft gels~\cite{shan95,CGS96,long96,kaji2013,karp15}. These phenomena play a role in a broad variety of applications, with many examples in the natural world and in technology. The equilibrium shapes of the drop and the elastic solid, and therefore also the contact angles, emerge from a balance between capillarity and elasticity \cite{L61,Rusanov75,Yuk86,Shanahan87,shan87b,White03,RomanJPCM,PBBB08,MPFPP10,SARLBB10,LM11,JXWD11,Style12,MDSA12b,Limat12,SBCWWD13,SchrollPRL13,Lubbers2014,Dervaux15,Bostwick2014,schu15}. 

%
\begin{figure}[b!]
\centerline{\includegraphics{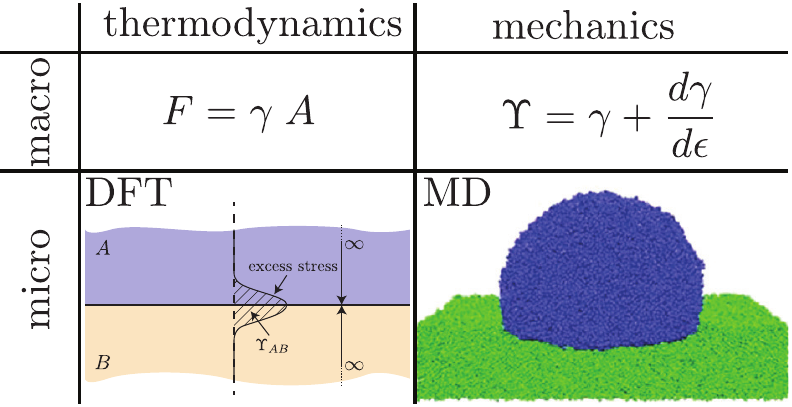}}
\vspace{-2 mm}
\caption{Perspectives on the Shuttleworth effect. Thermodynamics involves free energy minimisation, while the language of mechanics is expressed in terms of force balance. Macroscopically, these respectively involve the free energy per unit area $\gamma$, and the interfacial force per unit length $\Upsilon$. Microscopic equivalents, describing the molecular scale are given by Density Functional Theory (DFT) and Molecular Dynamics (MD, image from \cite{MDNeumann}).}
\vspace{-2 mm}
\label{fig:crossroads}
\end{figure}

Elastic interfaces exhibit an intriguing feature that is not present for liquid interfaces: the excess mechanical tension inside the interfacial region, referred to as the \emph{surface stress} $\Upsilon$, is in general different from the \emph{surface free energy} $\gamma$. This was pointed out already by Shuttleworth~\cite{Shuttleworth50} and studied in detail in crystals~ \cite{MS04}, with consequences in phenomena such as elastic instabilities \cite{MS04}, surface segregation \cite{WK77}, surface adsorption \cite{I04,grosman08}, surface reconstruction \cite{BGIE97,FF97}, nanostructuration \cite{R02} or  self-assembly \cite{AVMJ88,MSM05}. However, the consequences of $\Upsilon\neq \gamma$ in soft condensed matter are largely unknown \cite{Leiden2016}.
%
\begin{figure*}[t!]
\centerline{\includegraphics{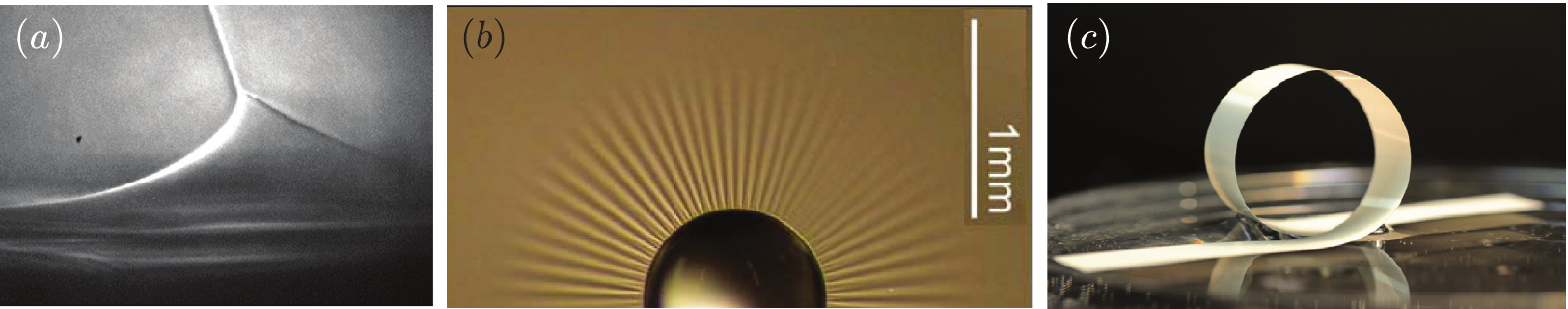}}
\vspace{-2 mm}
\caption{Elasto-wetting experiments. (a) X-ray visualisation of the deformation of a gel by a liquid below the contact line. Image from \cite{PARKNATURE} (b) Radial wrinkles induced on a floating thin membrane by a drop. Image from \cite{Huang2007}. (c) Elasto-capillary loop of a thin plate. Image from \cite{RomanJPCM}.}
\vspace{-2 mm}
\label{fig:pictures}
\end{figure*}

In this Perspective article we analyse the \emph{Shuttleworth effect} from different viewpoints, and unify thermodynamic and mechanical approaches (Fig.~\ref{fig:crossroads}). We discuss the conditions under which it influences wetting of deformable media (Fig.~\ref{fig:pictures}), and point to open questions.  


\section{Hierarchy of length scales}
\begin{figure*}[ht!]
\centerline{\includegraphics{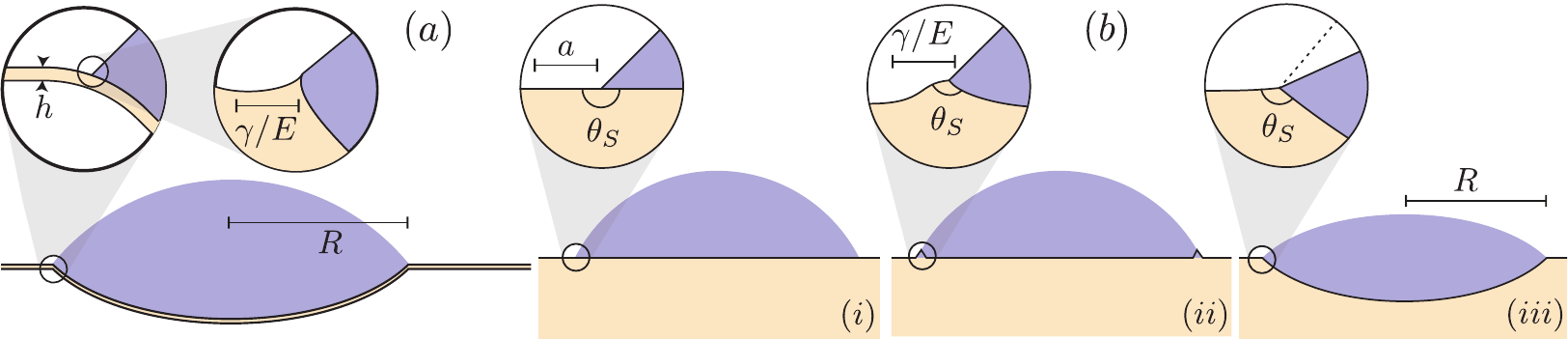}}
\vspace{-2 mm}
\caption{Scales of elasto-capillarity. (a) Drop of size $R$ on a thin membrane of thickness $h$. The first zoom shows that the sharp membrane bending is smooth on the scale $(B/\gamma)^{1/2}$. The second zoom shows wetting ridge on the scale $\gamma/E$. (b) Drops on a thick elastic layer upon varying $\gamma/E$. The three panels reveals the double transition of contact angles (microscopic versus macroscopic). The length scales evolve from $\gamma/E \ll a$ [panel (i)], to $a \ll \gamma/E \ll R$ [panel (ii)], to $R \ll \gamma/E$ [panel (iii)].}
\vspace{-2 mm}
\label{fig:scales}
\end{figure*}

Before discussing the Shuttleworth effect, it is important to assess the various regimes of elasto-capillarity. Such a classification can be made in terms of the relevant length scales \cite{RomanJPCM,SchrollPRL13}. Consider slender elastic bodies, whose thickness $h$ is much smaller than both its radius of curvature $\kappa^{-1}$ and its length $L$, such as that shown in Fig.~\ref{fig:scales}a, depicting a drop of size $R$ supported by a membrane of thickness $h \ll R$. Viewed at the scale of the drop, the thin membrane deforms sharply near the edge of the drop, forming well-defined contact angles. Zooming in near the contact line, however, the membrane angle varies gradually. Owing to the membrane's finite bending rigidity $B\sim Eh^3$, where $E$ is the Young's modulus, the bending occurs over a typical distance $\kappa^{-1} \sim (B/\gamma)^{1/2}$. This length is referred to as the \emph{bending-elasto-capillary length}. It provides, for instance, the characteristic size of the loop shown in Fig.~\ref{fig:pictures}c. However, the subsequent zoom in Fig.~\ref{fig:scales}a reveals a second length: the \emph{stretching-elasto-capillary length $\gamma/E$}. This is the scale over which the elastic solid deforms into a ``wetting ridge" in the direct vicinity of the contact line. 

The importance of the stretching-elasto-capillary length $\gamma/E$ becomes apparent when the elastic body is not slender. This is further highlighted for drops on very soft elastomers or gels (Fig.~\ref{fig:scales}b). Here one needs to introduce the range of molecular interactions $a$ as yet another length scale, setting the microscopic width of the interface. The three sequences of Fig.~\ref{fig:scales}b show a double transition of the contact angles~\cite{Lubbers2014,Dervaux15}. First, the microscopic contact angles change when $\gamma/E \sim a$ [from panel (i) to (ii)], without affecting the apparent angle on the scale of the drop. The macroscopic angles only change when $\gamma/E \sim R$ [from panel (ii) to (iii)]. These two transitions can be viewed as changes from Young's law to Neumann's law, respectively for the microscopic and macroscopic angles~\cite{Style12,MDSA12b,Limat12,Lubbers2014,Dervaux15}. It should be noted, however, that there still is no macroscopic derivation of Neumann's law for elastic substrates that includes the Shuttleworth effect. 

\section{The Shuttleworth equation} The surface energy $\gamma$ is the excess free energy per unit area of an interface. An area $A$ is therefore associated with an energy $\gamma A$. Applying the virtual work principle to an increase in interfacial area $\delta A$, one deduces the excess force per unit length $\Upsilon$. For a liquid interface, this leads to an increase $\gamma \delta A$ of the surface free energy. Equating this change in energy to the mechanical work done by $\Upsilon$, one obtains the identity $\Upsilon=\gamma$. For liquids there is thus no need to distinguish between  $\Upsilon$ and $\gamma$ and one simply refers to \emph{surface tension}. For elastic interfaces, however, the situation is fundamentally altered: expansion-induced strain changes the molecular structure of the interface. Hence, the interfacial excess free energy $\gamma$ is not constant any longer, and we find a change in free energy
\begin{equation}\label{eq:s1}
\delta( \gamma A)= \left(\gamma + A \frac{d\gamma}{dA}\right) \delta A = 
\left(\gamma + \frac{d\gamma}{d\epsilon}\right)\delta A.
\end{equation}
where $\epsilon$ is the strain parallel to the interface. Equating this to the work done by the surface stress, we find the Shuttleworth relation~\cite{Shuttleworth50,MS04}:
\begin{equation}\label{eq:shuttleworth}
\Upsilon =  \gamma + \frac{d \gamma}{d \epsilon}
\end{equation} 
In the context of soft matter, the derivative is understood as taken at constant chemical potential and temperature. The Shuttleworth effect gives rise to new phenomena that have no counterpart in liquids, whenever the surface free energy exhibits an explicit dependence on the strain.  

\section{Thermodynamics: Measuring the Shuttleworth effect}
\begin{figure}[b!]
\centerline{\includegraphics{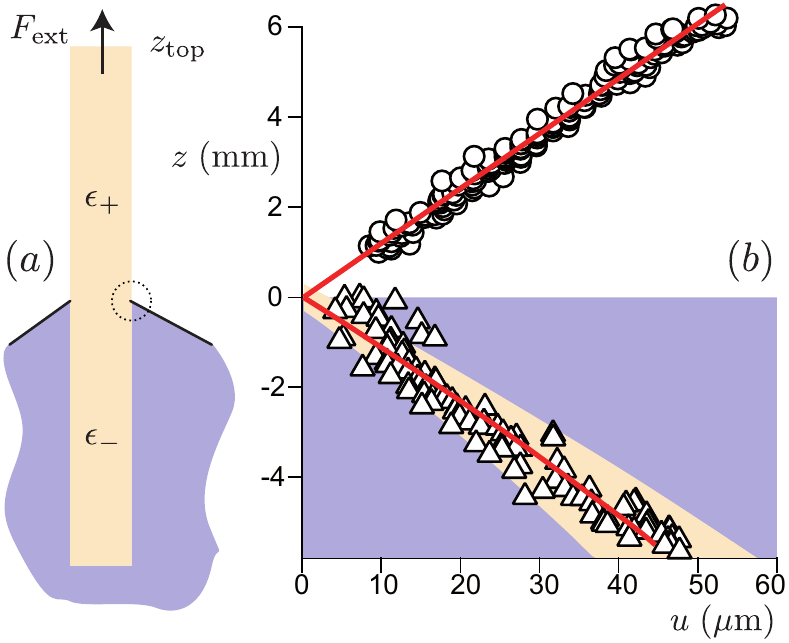}}
\vspace{-2 mm}
\caption{The elastic Wilhelmy plate: A tool to measure the Shuttleworth effect~\cite{MDSA12,Weijs2013}. (a) Schematic of an extensible plate or rod partially immersed in liquid bath, held by an external force. The Shuttleworth effect induces a discontinuity of strain across the contact line ($\epsilon_+ \neq \epsilon_-$). The small circle around the contact line is used in the free body diagram of Fig.~\ref{fig:young}. (b) Experimental measurement of the vertical displacement $u(z)$ along an elastic rod ($\gamma/E \sim 1\,\mu$m, radius $=150\,\mu$m). The discontinuity of strain $\epsilon=du/dz$ is clearly visible: the top part of the rod is stretched, the bottom part is compressed. Data from~\cite{MDSA12}.}
\vspace{-2 mm}
\label{fig:wire}
\end{figure}

The first illustration of the Shuttleworth effect is provided in a macroscopic thermodynamic framework. A suitable geometry to measure experimentally the strain derivative $\gamma' = d\gamma/d\epsilon$ consists of a slender, elastic plate or rod partially immersed in a liquid reservoir (Fig.~\ref{fig:wire}a). This setup forms an elastic realisation of the classical Wilhelmy plate~\cite{bookDeGennes}, normally used to measure liquid surface tension. As mentioned, the slender theory is derived under the assumption of a hierarchy of lengthscales
\begin{equation}\label{eq:wirehierarchy}
\mathrm{max}\left( \gamma/E, a \right) \ll h \ll \left(\frac{B}{\gamma} \right)^{1/2}.
\end{equation}
In this asymptotic limit, the spatial extent of the wetting ridge $\gamma/E$ is confined to a small region near the contact line, and its effect on the global energy of the plate is negligible. Hence, we are in the same hierarchy of scales as for intermediate zoom of the membrane in Fig.~\ref{fig:scales}a: the conclusions thus equally apply to the membrane, even though we consider the plate in Fig.~\ref{fig:wire}a under conditions where it remains straight~\cite{AndreottiPRE11}.

The strain away from the contact line (distances $\gg h$) is homogeneous, though we need to distinguish the strain in dry part of the plate ($\epsilon_+$) and the immersed part ($\epsilon_-$) -- cf Fig.~\ref{fig:wire}a. The macroscopic free energy reads~\cite{Weijs2013,Neuk2014}:
\begin{eqnarray}\label{eq:energywire}
\mathcal{E} &=& \gamma_{LV} A_{LV}  + \int_{\rm wet} dA \, \left[ 2\gamma_{SL}(\epsilon_-) + \frac{1}{2} Eh \epsilon_-^2  \right] \nonumber \\
 && - F_{\rm ext} z_{\rm top} + \int_{\rm dry} dA \, \left[ 2\gamma_{SV}(\epsilon_+) + \frac{1}{2} Eh \epsilon_+^2 \right].
\end{eqnarray}
Here $A_{LV}$ is the liquid-vapor area, while the solid-liquid and solid-vapor areas are represented by integrals over the ``wet" and ``dry" part of the surface --~factors $2$ reflect the two sides of the plate. Importantly, we allow for a dependence of the solid surface tensions on the strain $\epsilon$: this will give rise to the Shuttleworth effect. Other terms represent the bulk elasticity, as well as the work done by the external force $F_{\rm ext}$. We neglect bulk swelling so that the reference state is well defined.

The elasto-capillary equilibrium of the wire is obtained by minimisation of the free energy \cite{Weijs2013,Neuk2014}. In the Methods section, we briefly summarise the key steps of the derivation which can be performed for small strain, given the hierarchy of length scales. Variations of the contact line position and of $z_{\rm top}$, respectively, give two classical relations: Young's law for the contact angle $\theta$, and the force on the Wilhelmy plate, $F_{\rm ext} \sim \gamma_{SV}-\gamma_{SL} = \gamma_{LV} \cos \theta$. Variations with respect to $\epsilon_{+}$ and $\epsilon_{-}$ involve the derivative of the surface energy $\gamma'$ and lead to a new elasto-capillary coupling (see Methods):
\begin{eqnarray}
\epsilon_+ &=& \frac{2(\gamma_{SV}-\gamma_{SL})}{Eh}= \frac{2\gamma_{LV} \cos \theta}{Eh} \label{eq:upper}\\
\epsilon_- &=& \frac{2(\Upsilon_{SV}-\Upsilon_{SL})}{Eh}  \label{eq:lower},
\end{eqnarray}
where the reference state is the plate surrounded by air. 

Remarkably, the upper part of the plate probes the surface energies, while the immersed part probes the surface stresses. As a consequence, there is an elastic singularity in the form of a \emph{strain discontinuity} when crossing the contact line region \cite{Weijs2013,Neuk2014},
\begin{eqnarray}\label{eq:discontinuity}
\Delta \epsilon = \epsilon_+ - \epsilon_- = \frac{2(\gamma_{SL}'-\gamma_{SV}')}{Eh},
\end{eqnarray}
which directly quantifies the Shuttleworth effect. 

The discontinuity in strain has indeed been measured experimentally on a thin elastomeric rod. Figure~\ref{fig:wire}b shows the vertical displacement field measured on a wire of polyvinylsiloxane partially immersed in ethanol~\cite{MDSA12}. With respect to the reference state~---~the rod surrounded by air~---~the upper part of the rod is stretched ($\epsilon_+>0$), and the measured strain is in perfect agreement with the axisymmetric analogue of (\ref{eq:upper}). The lower part, however, is compressed ($\epsilon_-<0$). The discontinuity in strain can thus be used to quantify the strength of the Shuttleworth effect, which in the experiment gives~\cite{MDSA12}
\begin{equation}\label{eq:expvalue}
\gamma_{SL}'-\gamma_{SV}' =43\pm 10\,\mathrm{mN \,m^{-1}},
\end{equation} 
The magnitude of these terms is even larger than the relevant surface energies, $\gamma_{LV}=22.8\pm 0.2\,\mathrm{mN \,m^{-1}}$ and $\gamma_{SV}-\gamma_{SL} =16\pm 4 \,\mathrm{mN \,m^{-1}}$. Hence, for this material, the influence of the Shuttleworth effect in elasto-capillarity cannot be considered a small correction. 

\section{Macroscopic force balance near the contact line}
\begin{figure}[ht!]
\centerline{\includegraphics{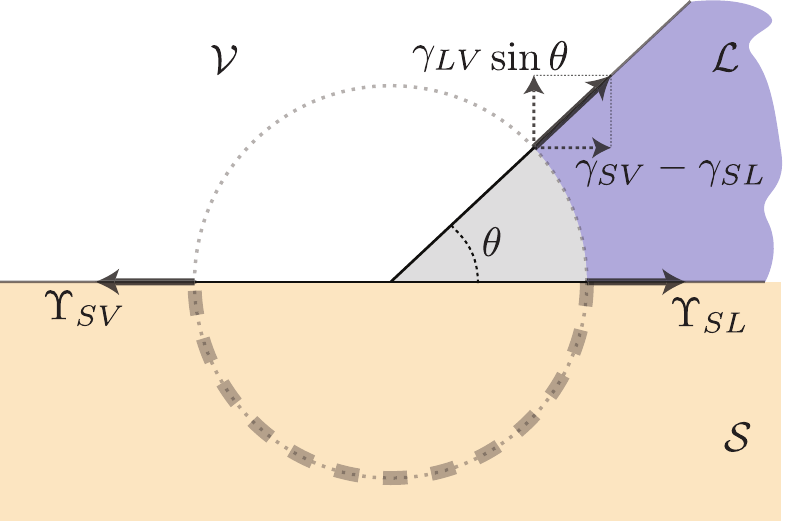}}
\vspace{-2 mm}
\caption{Macroscopic force balance near the contact line. We consider a mesoscopic region around the contact line of size $w \gg \gamma/E$. On the scale $w$, the solid interface remains flat and Young's law for the contact angle applies. The imbalance of surface stresses must be compensated by elastic stress on the lower boundary of the control volume (dashed line). The resultant elastic forces are $F_{\rm el}^n=\gamma_{LV} \sin \theta$ (normal) and $F_{\rm el}^t = \gamma'_{SL}-\gamma'_{SV}$(tangential). Adapted from \cite{Weijs2013}.}
\vspace{-2 mm}
\label{fig:young}
\end{figure}

We now turn to a mechanical view on the Shuttleworth effect. In the ``slender body" description of the extensible rod, the strain discontinuity (\ref{eq:discontinuity}) implies a perfectly localised line force of magnitude $\gamma_{SL}'-\gamma_{SV}'$, near each of the two contact lines. However, this slender formulation does not reveal the mechanics on the scale of the thickness $h$, let alone on the scales $a$ and $\gamma/E$. To gain insight in the force balance near the contact line, we now define a control volume of ``mesoscopic" size $w$ around the contact line. This control volume is indicated as the circle in Fig.~\ref{fig:wire}a and further detailed in Fig.~\ref{fig:young}. Its size $w$ is taken according to hierarchy of scales
\begin{equation}
\mathrm{max}\left( \gamma/E, a \right) \ll w \ll h.
\end{equation}
The first inequality ensures that the substrate remains essentially flat when viewed on the scale $w$: this is important since  in this limit the macroscopic contact angle $\theta$ still obeys Young's law. 

%

The forces acting on the mesoscopic control volume are indicated in Fig.~\ref{fig:young}. The normal component $\gamma \sin \theta$ of the liquid-vapor surface tension must be balanced by the elastic stress integrated over the bottom contour of the control volume. Contrarily to the common assumption found in the literature, however, the mechanical equilibrium also requires a \emph{tangential} elastic stress. Namely, owing to Young's law we can write the tangential liquid-vapor contribution as $\gamma_{LV} \cos \theta=\gamma_{SV}-\gamma_{SL}$. Importantly, the interfacial contribution along the elastic solid are expressed in $\Upsilon_{SV}$ and $\Upsilon_{SL}$, which in general differ from $\gamma_{SV}$ and $\gamma_{SL}$. Therefore, the horizontal surface stresses do not balance and a tangential elastic contribution is needed~\cite{Weijs2013}:
\begin{equation}\label{eq:ft}
F_{\rm el}^t = \gamma_{SV}-\gamma_{SL} + \Upsilon_{SL} - \Upsilon_{SV}= \gamma'_{SL} - \gamma'_{SV}.
\end{equation}
Just like the normal force, $F_{\rm el}^t$ must be balanced by the integral of elastic stress exerted on the contour delineating the bottom of the control volume. We thus find that the mechanical equilibrium near the contact line is truly elasto-capillary in nature: it requires both interfacial and bulk elastic contributions in normal and tangential directions.

\section{Thermodynamics versus mechanics}
We wish to emphasise that the ``mechanical" result (\ref{eq:ft}) is fully consistent with the ``thermodynamic" strain discontinuity  (\ref{eq:discontinuity}). The strain discontinuity that appears on the scale of the elastic rod ($\gg h$, Fig.~\ref{fig:wire}a) can be attributed to a tangential force $F_{\rm el}^t$ generated near the contact line ($\ll h$, Fig.~\ref{fig:young}). Only when there is no Shuttleworth effect, one finds $F_{\rm el}^t=0$ and a continuous strain across the contact line. 
The very same conclusion was recently drawn for a drop on a membrane  \cite{Hui2015b}: the Shuttleworth-induced discontinuity of strain implies a jump in the membrane tension across the contact line, altering the contact angles on the scale of the drop (Fig.~\ref{fig:scales}a).

\section{Microscopic origin of the Shuttleworth effect}
The experimental evidence of a significant Shuttleworth effect in a cross-linked polymer network (Fig.~\ref{fig:wire}b) comes as a surprise. Namely, one would have expected the structure of the surface at atomic scale to be close to that of an incompressible liquid, for which $\Upsilon=\gamma$. By contrast, for hard crystalline materials, the microscopic physics of the Shuttleworth effect is more easily understood~\cite{MS04}, e.g. from a toy model consisting of a network of masses and springs. Due to redistribution of electronic charge in the vicinity of the surface, the effective properties of the springs (rest length, spring constant) are different in the surface from the bulk. Then, the interfacial zone naturally exhibits an excess elastic stress $\Upsilon \neq \gamma$. 

\begin{figure}[ht!]
\centerline{\includegraphics{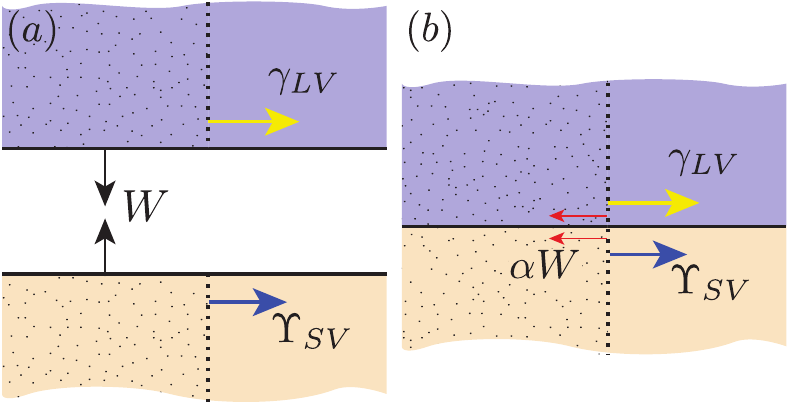}}
\vspace{-2 mm}
\caption{Microscopic view of the Shuttleworth effect, based on an approximate DFT model~\cite{Weijs2014}. (a) A liquid and solid phase are initially separated by a large vacuum. The surface stress is computed as the excess force due to the missing molecular interactions, exerted on the shaded region to the left of the dashed line. (b) Joining the liquid and solid phases releases (per unit area) the work of adhesion $W=\gamma_{LV}+\gamma_{SV}-\gamma_{SL}$. The vertical attraction leads to a compression of the interfacial zone: the Shuttleworth effect arises whenever a fraction $\alpha <1$ is transmitted horizontally.}
\vspace{-2 mm}
\label{fig:micro}
\end{figure}

How can we understand the Shuttleworth effect for the liquid-like molecular structure of a cross-linked polymer network? In the continuum framework of Density Functional Theory (DFT), in the sharp interface approximation, it has been been possible to relate the Shuttleworth effect to a \emph{compressibility of the interfacial layer}~\cite{Weijs2014}. The corresponding physics is summarised in Fig.~\ref{fig:micro}. Panel (a) shows a liquid phase (top) and an elastic phase (bottom) that are separated by a large vapour layer that can effectively be treated as a vacuum. The respective surface stresses are $\Upsilon_{LV}=\gamma_{LV}$ and $\Upsilon_{SV}=\gamma_{SV}+\gamma_{SV}'$. Panel (b) shows that bringing the two interfaces together leads to a vertical attractive interaction, with a strength given by the work of adhesion:
\begin{equation}\label{eq:wadh}
W=\gamma_{LV}+\gamma_{SV}-\gamma_{SL}=\gamma_{LV}(1+\cos \theta).
\end{equation}
After joining the solid-liquid phases, this attraction leads to an extra compressive stress in the interfacial zone  in the normal direction. For an incompressible layer this extra normal compression is equally transmitted in the tangential direction. However, this is not the case when the interfacial layer is compressible: only a fraction $\alpha W$ of this extra stress is retransmitted in the tangential direction (Fig.~\ref{fig:micro}b, red arrow). One can express this fraction $\alpha=\nu_s/(1-\nu_s)$ in terms of an interfacial Poisson ratio $\nu_s$. Summing the various contributions depicted in Fig.~\ref{fig:micro}b gives the solid-liquid surface stress $\Upsilon_{SL} = \Upsilon_{SV} + \gamma_{LV}  - \alpha W$. Hence, using (\ref{eq:shuttleworth}) and (\ref{eq:wadh}), the microscopic model predicts
\begin{equation}\label{eq:alpha}
\gamma_{SL}'-\gamma_{SV}'=(1-\alpha)W.
\end{equation} 

%

\begin{figure}[ht!]
\centerline{\includegraphics{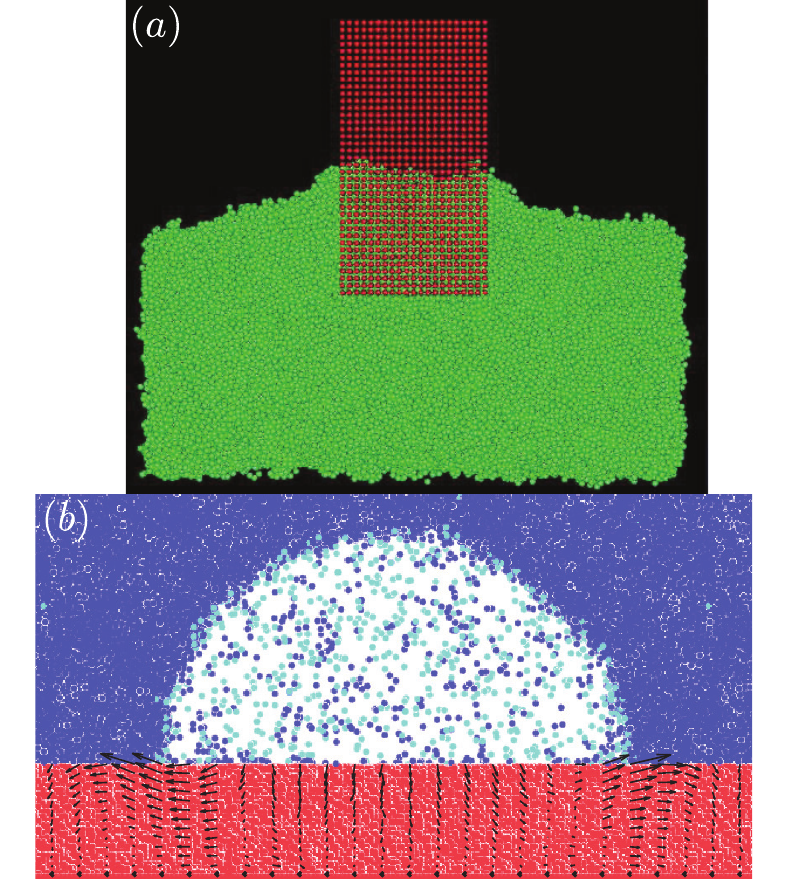}}
\vspace{-2 mm}
\caption{The Shuttleworth effect in Molecular Dynamics. (a) A rigid probe partially immersed in a Lennard-Jones liquid. The simulations measure a large tangential force $\approx W$ on the solid near the contact line. Image from~\cite{SevenoPRL13}. (b) A gas bubble in a Lennard-Jones liquid on a deformable elastic solid, in a case where $\gamma_{SL}= \gamma_{SV}$. The black arrows show the displacement field inside the solid: despite the symmetry in surface energies, the displacements are biased towards the liquid side due to $\Upsilon_{SL}\neq \Upsilon_{SV}$. Data from~\cite{Weijs2013}}\label{fig:MD}
\vspace{-2 mm}
\end{figure}

While the DFT model represents a highly simplified description of the molecular structure at the interface, Eq.~(\ref{eq:alpha}) provides two important predictions~\cite{Weijs2014}: 
\begin{itemize}
\item the departure from liquid-like behaviour relates to the compressibility of the interfacial zone,
\item the work of adhesion gives an upper bound on the Shuttleworth effect ($\alpha=0$):
\begin{equation}\label{eq:max}
F_{\rm el}^t = \gamma_{SL}'-\gamma_{SV}' < W = \gamma_{LV}(1+\cos \theta).
\end{equation}
\end{itemize}
The importance of the work of adhesion has been confirmed quantitatively in Molecular Dynamics (MD) simulations. Figure~\ref{fig:MD}a shows a simulation of a rigid Wilhelmy plate, intended as a model for an AFM tip~\cite{SevenoPRL13}. The simulations measured the total force that the liquid molecules exert on the solid near the contact line: a large tangential force component was found, with a strength consistent with the work of adhesion $\gamma_{LV}(1+\cos \theta)$~\cite{SevenoPRL13}, as predicted in~\cite{DMAS11}. The often claimed $\gamma_{LV} \cos \theta$ is clearly not observed. A similar MD simulation of a deformable Wilhelmy plate showed a strain discontinuity, which using (\ref{eq:discontinuity}) gave a Shuttleworth effect close to the upper bound set by the work of adhesion (\ref{eq:max})~\cite{Weijs2013}. Finally, also the experimental value for the elastomeric wire, quoted in (\ref{eq:expvalue}), is very close to the upper bound. 

A final striking demonstration of the Shuttleworth effect is given in Fig.~\ref{fig:MD}b. The picture represents a snapshot of a bubble on a weakly deformable wall, with the hierarchy of length scales $\gamma/E \ll a\ll R$. The contact angle is close to $90^\circ$, which implies $\gamma_{SL}\approx \gamma_{SV}$. Despite this symmetry in surface energies, the elastic deformation is very asymmetric~\cite{Weijs2013}: the black arrows in Fig.~\ref{fig:MD}b represent the displacement field inside the solid, clearly showing a strong tangential displacement towards the exterior of the bubble. The breaking of symmetry is due to $\Upsilon_{SL} \neq \Upsilon_{SV}$, even though $\gamma_{SL}\approx \gamma_{SV}$. The bias towards the liquid side is perfectly in line with $F_{\rm el}^t = \gamma_{SL}' - \gamma_{SV}' \sim W$.

\section{Perspective} 
Interfacial effects of soft solids provide an opening playground in soft condensed matter \cite{Leiden2016}. We have presented here some fundamental aspects of the coupling between elasticity and capillarity --~focusing on the so-called Shuttleworth effect that arises when surface free energy depends explicitly on the elastic strain. While we focussed on the prototypical ``liquid drop on elastic solid", the same issues and subtleties arise for adhesion of very soft solids~\cite{styl13c,sale13,Cao2014,HuiProc2015,KarpSM2016}.    

Figure~\ref{fig:wire} exemplifies a system (polyvinylsiloxane in contact with ethanol) that presents a strong strain-dependence $d\gamma/d\epsilon$. By contrast, no Shuttleworth effect was needed to accurately describe the wetting behaviour of water on polydimethylsiloxane~\cite{SBCWWD13,Bostwick2014}. A key open issue is therefore to understand the physicochemical conditions for the appearance of the Shuttleworth effect. Is it important that the liquid is a good rather than a bad solvent, as suggested by the interpretation in terms of surface compressibility? A more detailed understanding of the interface of a reticulated polymer with a liquid, and its consequences for elasto-capillary mechanics, is necessary. For example, we have ignored here the distance between crosslinks or entanglement points, which determines the scale at which entropy dominated elasticity and reference state are defined in the continuum. There is yet another length scale, associated with distinction between bulk swelling \cite{Kajiya2013,Dupas2014} and interfacial effects associated with polymer free ends at the free surface of the sample. There is an urgent need for a systematic, quantitative characterisation of polymers and liquids, using a reliable set-up to measure the Shuttleworth effect. Can one design an alternative method to the experiment shown in Fig.~\ref{fig:wire}, which required a very high resolution? The task is obviously difficult, as it directly involves stretching \cite{NadermannPNAS2013,MondalPNAS2015}: bending and buckling effects, which nicely lead to amplified elastic effects, are completely decoupled from the Shuttleworth effect.

We have highlighted here several intriguing consequences of the Shuttleworth effect. However, the detailed examples given in this paper relied on the stretching-elasto-capillary length $\gamma/E$ being relatively small. A systematic, fully consistent analysis still remains to be done for the more interesting cases of highly deformed interfaces. To give a striking example, even the selection of the contact angle for a droplet on a soft gel in the presence of the Shuttleworth effect is still an open problem on which contradictory statements can be found in the literature. Undoubtedly, soft elastic interfaces will continue to stretch our intuition for capillarity. 

\section{Methods}
The derivation of the strain discontinuity (\ref{eq:discontinuity}) involves some subtle kinematics: the contact line position $z_{\rm cl}$ and the top of the plate $z_{\rm top}$ are not independent of the strains $\epsilon_\pm$. Here we briefly sketch the essential steps, which are properly developed in~\cite{Weijs2013,Neuk2014}. First, we remind that the external force (per unit plate width) is $F_{\rm ext}=2(\gamma_{SV}-\gamma_{SL})$, which can be derived from a vertical displacement $\delta z_{\rm top}$ at constant $\epsilon_\pm$. The factor 2 is due to the two sides of the plate. Next, we consider variations $\delta \epsilon_\pm$ while keeping the contact line position fixed. The lengths of the dry/wet parts can be written as $L_\pm(1+\epsilon_\pm)$, where $L_\pm$ are lengths in the reference state. Expressing $z_{\rm top}=z_{\rm cl} + L_+(1+\epsilon_+)$, the energy per unit width (\ref{eq:energywire}) becomes
\begin{eqnarray}\label{eq:energywirebis}
\mathcal{E} &=& L_+ (1+\epsilon_+) \left[ 2\gamma_{SV}(\epsilon_+) + \frac{1}{2} Eh \epsilon_+^2 - F_{\rm ext} \right] \nonumber \\
 &+& L_- (1+\epsilon_-) \left[ 2\gamma_{SL}(\epsilon_-) + \frac{1}{2} Eh \epsilon_-^2  \right],
\end{eqnarray}
where we omitted redundant constants. Minimisation now reduces to $\partial \mathcal{E}/\partial \epsilon_\pm=0$, which gives the equilibrium conditions (for $|\epsilon_\pm| \ll 1$):
\begin{eqnarray}
2(\gamma_{SV}+ \gamma_{SV}') + Eh \epsilon_+ &=&  F_{\rm ext} = 2(\gamma_{SV} - \gamma_{SL}) \nonumber \\
2(\gamma_{SL} + \gamma_{SL}') + Eh \epsilon_- &=&0.
\end{eqnarray}
The combination $\gamma+\gamma'$ emerges in the same manner as in (\ref{eq:s1}), where we derived the Shuttleworth equation for the surface stress. Combining the two equilibrium conditions yields the strain discontinuity (\ref{eq:discontinuity}). Equations (\ref{eq:upper},\ref{eq:lower}) are obtained after subtracting the reference strain in air $\epsilon_0=-(\gamma_{SV}+\gamma_{SV}')=-\Upsilon_{SV}$.


\section{Acknowlegments}
We are indebted to our collaborators, in particular Antonin Marchand, Joost Weijs, Siddhartha Das, Hugo Perrin, Romain Lhermerout, Anupam Pandey and Stefan Karpitschka. We are also grateful to the participants of the Lorentz Center workshop ``Capillarity of Soft Interfaces" for stimulating discussions that led to this perspective article. Financial support is acknowledged from the 3TU Center of Competence (Fluid and Solid Mechanics), ERC (the European Research Council) Consolidator Grant No. 616918 and ANR SMART.


\vspace{-2 mm}

\end{document}